# Massive liquid Ar and Xe detectors for direct Dark Matter searches

B.M. Ovchinnikov[1], Yu.B. Ovchinnikov[2], V.V. Parusov[1]

[1]Institute for Nuclear Research of Russian Academy of Sciences, Moscow, Russia

[2]National Physical Laboratory, Teddington, Middlesex, TW11 0LW, UK

**Abstract**

A novel experiment for direct searches of the Dark Matter with liquid argon double-phase chamber with a mass of liquid Ar up to several hundred tons is proposed. To suppress the β-, γ- and $n_0$- backgrounds, the comparison of scintillation and ionization signals for every event is suggested. The addition in liquid Ar of photosensitive $Ge(CH_3)_4$ or $C_2H_4$ and suppression of triplet component of scintillation signals ensures the detection of scintillation signals with high efficiency and provides a complete suppression of the electron background. For the detection of photoelectrons and ionization electrons, highly stable and reliable GEM detectors must be used.

## 1. Introduction

The experiment for direct searches of Weakly Interacting Massive Particles (WIMPs) with a 100 liter Xe (Ar) double-phase chamber was proposed in [1]. The addition of trimethylamine (TMA) for detection of scintillation signals with efficiency of 35% was suggested earlier in [2]. As a result of absorption of the scintillation light, emitted by the atoms and molecules of Ar, excited in a process of recoil atom and background electron interaction, the ionization core is surrounded by a cloud of photoelectrons, the diameter of which is determined by the concentration of TMA. The detection of the ionization core and photoelectrons, performed with high spatial resolution and efficiency, makes it possible to compare the ionization and scintillation signals and to highly suppress the background events.

Using of TMA as a photosensitive addition for the chamber with a large mass of liquid Ar is not possible because of slight electro negativity of the TMA molecules [2-7]. Due to that fact, the drift length of electrons in liquid Xe+40 ppm TMA, at the field of $3kV\cdot cm^{-1}$, is as small as 10 cm [1].

The first results on Dark Matter searches with a 2.3-litre liquid Ar chamber have been achieved in [8], where the scintillation signals are detected by photomultipliers (PMT) with an efficiency of a few percent. All inner surfaces of the chamber are covered with tetra-phenyl-butadiene wave-shifter, which is used for conversion of the VUV light into visible emission. Electrons of the ionization core are transported with electric field from the liquid Ar to the gas phase, where they produce an electroluminescent light, which is detected by PMT.

The one ton liquid Ar chamber is constructed in [9], where GEM detectors, made of metalized kapton, are proposed for the detection of electrons from an ionization core. However, using of plastic GEM detectors in long-term experiments is rather problematic due to their low reliability [10-13].

In this work it is proposed to use $Ge(CH_3)_4$ (TMG) [14] or $C_2H_4$ [19] as a photosensitive additions to the liquid Ar of the detector. Since TMG and $C_2H_4$ are non-electronegative molecules, the electron drift length in the Ar +TMG ($C_2H_4$) mixtures may be equal to several meters, supposed that these mixtures can be purified from almost all electronegative impurities [15-18].

In liquid Ar the energy of photons $Ar_2^* \rightarrow 2Ar + h\nu$ is equal to ~(7.5-11.5) eV, while the ionization potential of TMG in liquid Ar is about 8.55eV and for $C_2H_4$ it is below 9.8 eV. As a result of that, the photons, emitted by $Ar_2^*$, can ionize TMG ($C_2H_4$) molecules with producing of a cloud of photoelectrons around the ionization core [14] (Fig.1).

Recoil ions, secondary and background electrons are ionizing Ar atoms by producing electron – hole pairs and exited atoms, excitons, with the following, in both cases, of self – trapped exciton luminescence, which consists of fast singlet $^1\Sigma_u^+$ component with $\tau_1$ =7±1 ns and slow triplet $^3\Sigma_u^+$ component with $\tau_2$ =1.6±0.1 $\mu s$ [8].

The ratio between the singlet and triplet intensities in liquid Ar [8] is equal to

$$I_s/I_t=0.3(e^-),\ 1.3(\alpha) \text{ and } 3(nr). \qquad (1)$$

For suppression of the electron background in some works (see Table 1) the criterion F is used:

$$F=I_s/(I_s + I_t). \qquad (2)$$

An alternative method for electron background suppression is a comparison of ionization $S_2$ and scintillation $S_1$ signals for every event. The ratio between the ionization and scintillation signals in liquid Ar for E=100keV, at the electric field intensity of 1 kV/cm, is equal to

$$S_2/S_1=150(e^-),\ 3(\alpha) \text{ and } 10(nr)\ [8]. \qquad (3)$$

The additional charge in liquid Ar + TMG(0.15ppm) is produced mainly by photons, because the probability of the ionization at the $Ar_2^*$+TMG-collisions at this concentration of TMG is small [7].

The quantum efficiency of photo-ionization $h\nu$(128 nm)+TMG$\rightarrow$ $TMG^+ + e^-$ in a liquid Ar is equal to 50% [19].

The ionization core of a recoil ion or a background electron, at the concentration of TMG of 0.15 ppm, is surrounded by a photoelectron cloud of ~10 cm in diameter [14].

The measurement of the number of photoelectrons, surrounding the ionization core, ensures the high efficiency measurement of the scintillation signal amplitude, the efficiency of which must be better than in a case of using of photomultipliers, and this should make it possible to obtain high suppression of the background.

The dependence $S_2/S_1=f(E_{nr})$ is shown in Fig.5 of work [8]. The quantity $S_2/S_1$, obtained by extrapolation, is equal to 40 for $E_{nr}$ =10keV. The number of photons for $E_{nr}$=30keV is equal to 300 [8]. The ratio $S_2/S_1$ for electron background in the working range is not very different.

At present several experimental installations with high mass of liquid Ar and Xe for Dark Matter search have been proposed and developed, the main parameters of which are given in Table 1.

## 2. Experimental setup

As an example, the design of a chamber with a mass of liquid Ar of 20 tons is shown in Fig.2. The ionization electrons and photoelectrons are detected with GEM of high reliability and stability [10-13]. The diameter of GEM is equal to ~70 cm.

The high three-dimensional spatial resolution of GEM ensures the detection of the ionization core and photoelectrons with high efficiency.

The light screen [24] is placed between the liquid Ar and the GEM detector to block the transmission of photons from the GEM to the volume of liquid Ar. On the other hand, this screen is transparent for electrons, which are passing through it from the liquid Ar to GEM with almost 100% efficiency. To suppress the secondary scintillation from electrons moving in the gas phase of Ar in the region below the light screen, the hydrogen under a partial pressure of about 0.1 bar must be added to Ar gas phase.

The events, which take place near the cathode and the surface of liquid Ar can be picked out by analysis of the shape of photoelectron clouds.

To reduce the background produced by the cathode, it must be winded with a carbon wire of ∅1 mm at 50 mm pitch.

To suppress the background, the 5 cm thick layer of copper can be placed within the chamber.

The ring electrodes for the electric field forming must be produced from mylar covered with copper layer of about 5μm thickness.

The neon under a partial pressure of ~1 bar [10-13] must be contained in the gas phase under the liquid Ar to decrease the difference of the potential between the GEM electrodes.

The earlier developed methods of purification of noble gases up to level of $10^{-11}$ rel. vol. parts $O_2$ [15-18] makes it possible to transport the ionization electrons, at the electric field intensity of 1kV/cm, over a distance of 10 meters.

The errors for scintillation and ionization signals measurements may be equal to 20%.

The addition of about 100ppm Xe($N_2$) in liquid Ar allows to suppress the triplet component of scintillation signals and to increase the coefficient of background suppression. The triplet component of scintillation signals was suppressed practically completely with addition in liquid Ar of about 100ppm Xe [25]. The singlet component is changed little for this addition. The relationships (3) in this case give the next numbers:

$$S_2/S_1=645(e^-),\ 5.31(\alpha) \text{ and } 13.3(nr). \qquad (4)$$

The relationship $(S_2/S_1)_e/(S_2/S_1)_{nr}=48.5$ allows to suppress the electron background completely.

The neutron background must be essentially suppressed due to multiple elastic and inelastic scattering of neutrons at Ar atoms in the course of their slowing [1]. The high 3D spatial resolution of the chamber allows to restore the picture of neutron scattering and to suppress this background too.

Since the decay rate of $Ar^{39}$ in 20 tons of natural argon is about $2\times10^4\ s^{-1}$, for the total collection time of electrons from the volume of the detector of 2 ms, only about 40 background events can take place in the volume of the 20 tons chamber during that time, which gives 1 event per 500 kg of Ar. Therefore, the probability of overlapping of two such events is small and the $Ar^{39}$ background can be completely suppressed. In a similar way the $Ar^{39}$ background can be suppressed for even larger detection chambers, with volume of 1000 tons, for example.

## References

1. B.M.Ovchinnikov, V.V.Parusov,"The suppression of background in experiment for WIMP search with double phase Ar(Xe) chamber", Preprint INR of RAS -0966/1997;
   "A method for background reduction in an experiment for WIMP search with a Xe (Ar) – liquid ionization chamber", J. Astroparticle Physics **10** (1999) 129.
2. S.Suzuki, T.Doke, A.Hitachi et al., "Photo ionization in liquid Ar doped with trimethylamine or triethylamine", NIM **A245** (1986) 366.
3. S.Kubota, A.Nakamoto, T.Takahashi et al., "Evidence of the existence of exciton states in liquid Ar and exciton-enhanced ionization from Xe doping", Phys. Rev. **B13**, No4 (1976) 1649.
4. D.F.Anderson, "Photosensitive dopants for liquid Ar", NIM **A242** (1986) 254.
5. D.F.Anderson, "New photosensitive dopants for liquid Ar", NIM **A245** (1986) 361.

6. S.Suzuki, T.Doke, A.Hitachi et al., "Photo ionization in liquid Xe doped with TMA or TEA", NIM **A245** (1986) 78.

7. S.Kubota, M.Hishida, M.Suzuki, J.Ruan, "Liquid and solid Ar, Kr and Xe scintillators", NIM **196** (1982) 101.

8. WARP collaboration, "First results from a Dark matter search with liquid Ar at 87 K in the Gran Sasso Underground Laboratory".

9. A.Marchionni, C.Amser, A.Badertscher et al., "Ar DM: A ton-scale liquid Ar detector for direct dark matter search", arXiv:1012.5967.

10. B.M.Ovchinnikov, V.V.Parusov, Yu.B.Ovchinnikov, "A multi-channel wire gas electron multiplier", arXiv:1003.1240; PTE No5 (2010) 37; Instr. and Exper. Techn., **53**, No 5 (2010) 653.

11. B.M.Ovchinnikov, V.V.Parusov, Yu.B.Ovchinnikov, "A multi-channel wire gas electron multipliers with gaps between the electrodes 1 and 3 mm", arXiv:1005.1556; PTE No6 (2010) 68; Instr. and Exper. Techn., **53**, No 6 (2010) 836.

12. B.M.Ovchinnikov, V.V.Parusov, Yu.B.Ovchinnikov, "A multi-channel gas electron multipliers with metallic electrodes", arXiv:1006.2986; PTE No6 (2010) 3.

13. B.M.Ovchinnikov, V.V.Parusov, Yu.B.Ovchinnikov, "The gas electron multipliers with high reliability and stability", arXiv:1012.4716.

14. P.Cennini, J.P.Revol, C,Rubbia et al.,"Improving the performance of the liquid Ar TPC by doping with tetra-methil-germanium", NIM **A355**(1995)660

15. A.S.Barabash, O.V.Kazachenko, A.A.Golubev, A.N.Zelensky, B.M.Ovchinnikov, "The system for deep Ar purification and device for electronegative admixtures measurements in noble gases and hydrocarbons", PTE, No6,(1978) 189.

16. A.S.Barabash, O.V.Kazachenko, A.A.Golubev, B.M.Ovchinnikov, "Catalytic and adsorbing purification of NG, $H_2$, $CH_4$ from $O_2$", J. "Chemical industry", 1984, No6, p.373.

17. A.S.Barabash, B.M.Ovchinnikov, "The purification of *Ar* for liquid ionization chambers with zeolites ", Preprint INR of RAS, П-0061, Moscow, 1977.

18. A.S.Barabash, A.A.Golubev, O.V.Kazachenko, B.M.Ovchinnikov, "Liquid pulsed ionization chamber, filled with Xe, Ar and $CH_4$", Preprint INR of RAS, 1984, П-034, Moscow; NIM **A236** (1985) 69.

19. V. Vuillemin, P.Cennini, C.W.Fabjan et al. "Electron drift velocity and characteristics of ionization of alpha and beta particles in liquid argon doped with ethylene for LNC calorimetry", Preprint CERN-PPE/91-171, 1 October 1991.

20. MiniClean Collaboration Los Alamos, "The MiniClean Dark Matter Experiment", Proceedings of the DPF-2011Conference, Providence, RI, August 8-13, 2011.

21. Laura Baudis, "Darvin: dark matter WIMP search with noble liquids", arXiv: 1012.4764V1 [astro-ph. IM] 21 Dec. 2010.

22. K. Arisaka, C.W. Lam, P.F. Smith at al., "Studies of a three-stage dark matter and neutrino observatory based on multi-ton combinations of liquid Xe and liquid Ar detectors", arXiv: 1107.1295V1 [astro-ph. Im] Jul. 2011.

23. B.M. Ovchinnikov, Yu. B. Ovchinnikov, V.V. Parusov, "Proposal on large-scale liquid Ar detector for direct Dark Matter searches", Preprint INR 1307/2011, November 2011; arXiv: 1102.3588, 2011.

24. B.M.Ovchinnikov, V.V.Parusov, T.A.Murashova, V.A.Popov. "A lightproof screen for time-projection chambers with large multiplication coefficients and photo cathodes", Instrum. Exp. Tech. **42** (1999) 386.

25. S.Kubota, M.Hishida, M.Susuki, J.Ruan, “Liquid and solid Ar, Kr andXe scintillators” NIM 196(1982)101.

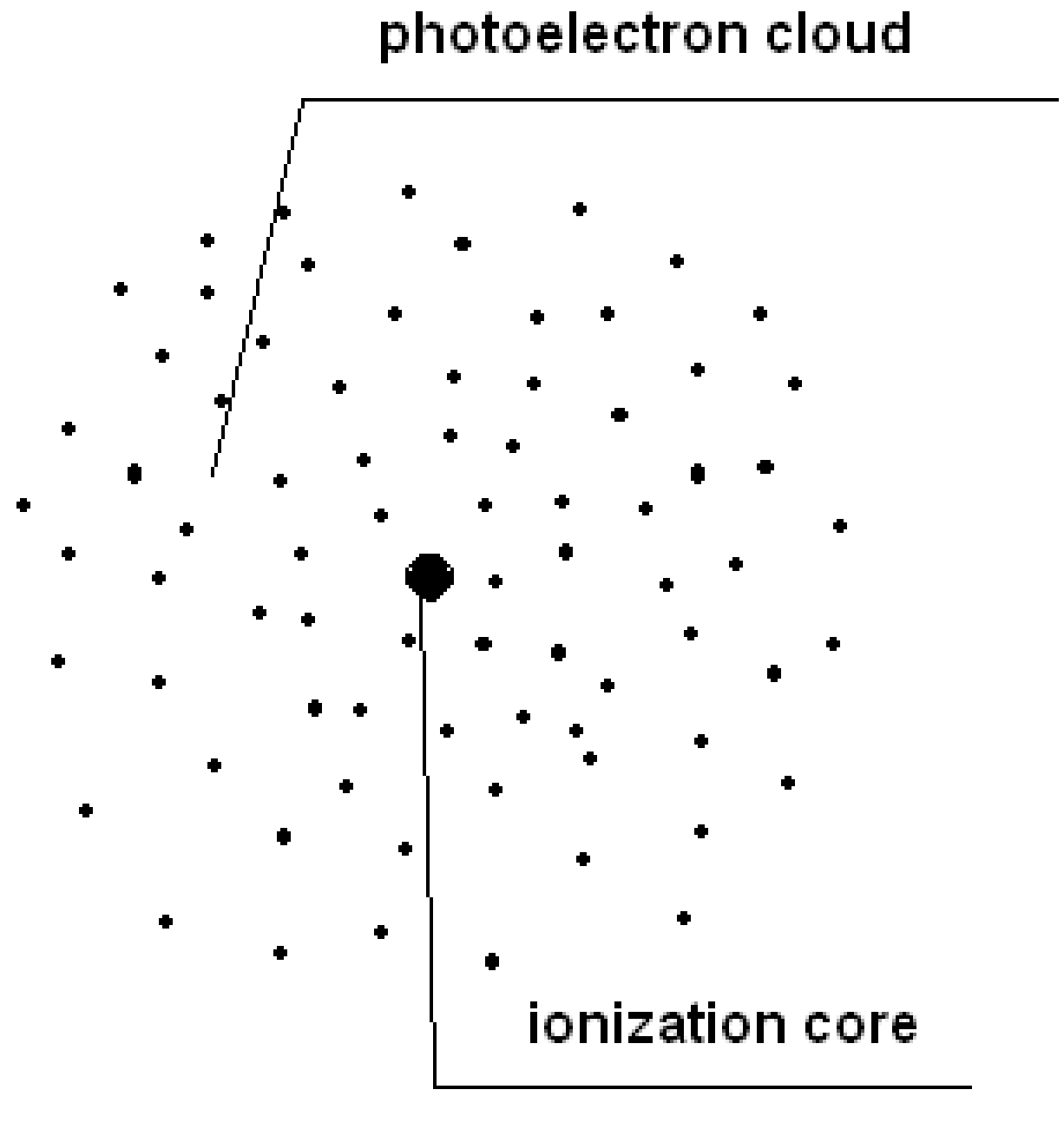

Fig 1.The event structure

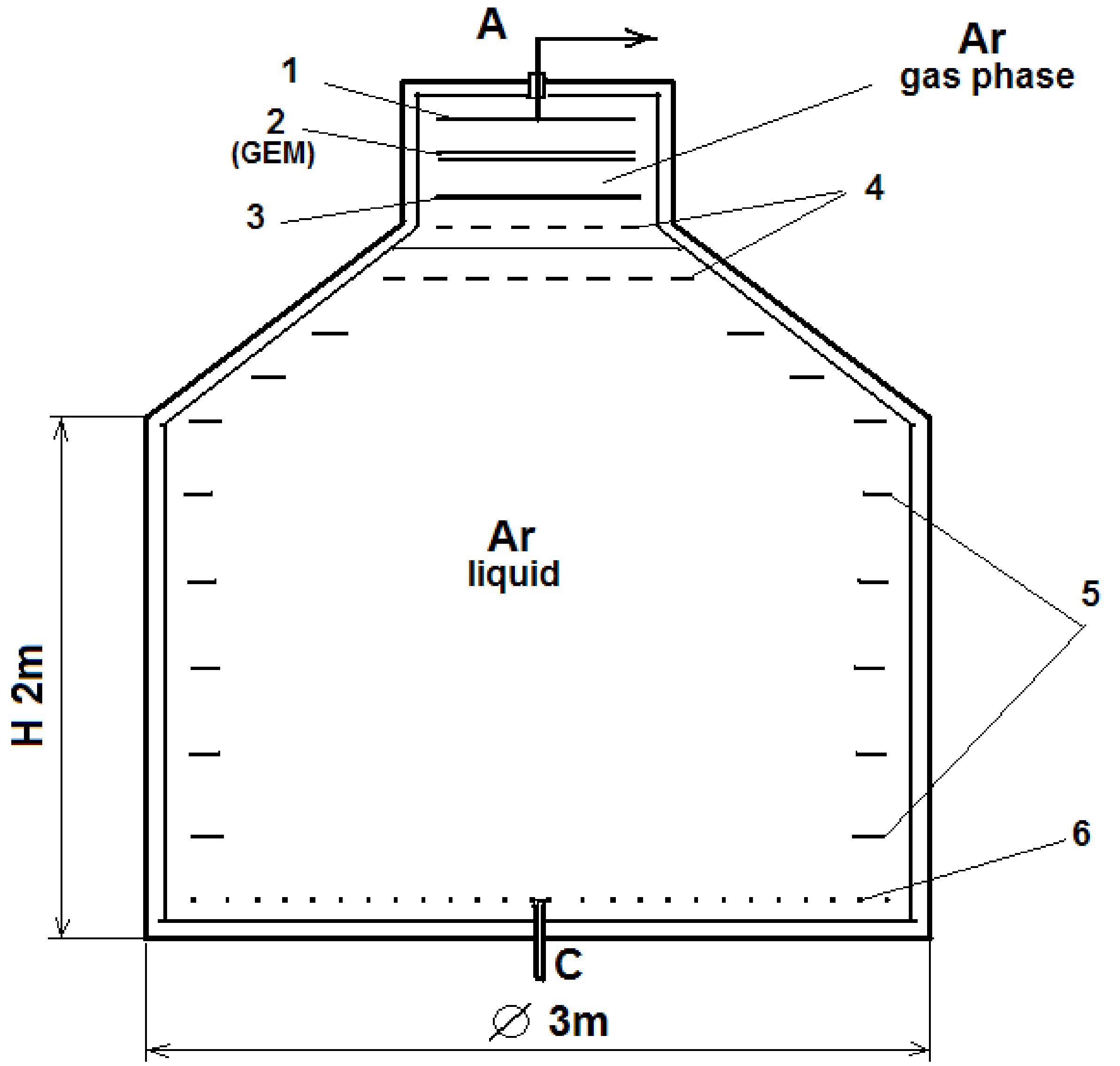

Fig.2. Double-phase Ar chamber: 1-anode, 2-GEM detectors, 3-light screen, 4- grids for electrons transport from liquid Ar to gas phase, 5-the ring electrodes for electric field level, 6-canhode.

Table 1.

| The name of project | The target of detector | The detection method | The threshold of detection | The method for background suppression | $Ar^{39}$ concentration in Ar | The expected result |
|---|---|---|---|---|---|---|
| "ArDM"<br>A.Rubbia [9] | Ar<br>1000(850) kg<br>double-phase | $S_2/S_1$+F<br>PMT+GEM | $E_{nr}^{min}$ =30 кэВ | $S_2/S_1$+F | $10^2$ decays/t·s | σ(WIMP)<br>$=10^{-45}$cm$^2$ |
| "MiniClean"<br>Los Alamos [20] | Ar liquid<br>500(150) kg<br>single-phase | F<br>92 PMT | $E_{nr}^{min}$ =30 кэВ<br>$M^{min}$(WIMP)<br>=20 GeV | F | $10^3$ decay/t·s | $10^{-45}$см$^2$ |
| "Deap-3600"<br>Los Alamos [20] | Ar liquid<br>3600(1000)kg<br>single-phase | F<br>266 PMT | | F | $10^3$ decays/t·s | $10^{-46}$ cm$^2$ |
| "Clean"<br>Los Alamos [20] | Ar liquid<br>40(10)tons<br>single-phase | F<br>PMT | $M^{min}$(WIMP)<br>=60 GeV | F | <$10^2$ decays/t·s | $6·10^{-47}$ cm$^2$ |
| "Darvin"<br>[21] | Ar<br>20(10)tons<br>double-phase | $S_2/S_1$+F<br>avalanche photodiodes<br>+GEM | $E_{nr}^{min}$ =30 кэВ | $S_2/S_1$+F<br>$К_e$ suppression=$10^8$ | <40 mBq/kg | $4·10^{-48}$ cm$^2$ |
| | Xe<br>8(5) tons<br>double-phase | F+$S_2/S_1$<br>avalanche photodiodes<br>+GEM | $E_{nr}^{min}$ =10 кэВ | $S_2/S_1$+F<br>К=$10^3$ | $10^{-4}$ decays/kg·day·keV | |
| Los Angeles<br>Dr. D.Cline [22].<br>(proposal) | Ar<br>580(500)tons<br>double-phase | $S_2/S_1$<br>12000 avalanche<br>photodiodes | $M^{min}$(WIMP)<br>≅20GeV | $S_2/S_1$+F | <10 decays/t·s | $10^{-48}$ cm$^2$ |
| | Xe<br>146(100)tons<br>double-phase | $S_2/S_1$+F<br>3740 avalanche<br>photodiodes | $M^{min}$(WIMP)<br>≅6 GeV | $S_2/S_1$+F<br>К=$10^3$ | | |
| INR of RAS<br>(proposal) [23] | Ar+$Ge(CH_3)_4$<br>(0,15 ppm) or<br>Ar+$C_2H_4$ (2 ppm)<br>1000tons<br>double-phase | $S_2/S_1$<br>GEM | $E_{nr}^{min}$ =10 кэВ | $S_2/S_1$<br>$К_e$ >$10^{10}$ | $10^3$decays/t·s | ≤$10^{-48}$ cm$^2$ |